\documentclass[]{aa}
\usepackage[varg]{txfonts}

\usepackage{graphicx,rotating}
\usepackage{amssymb,amsmath,pdflscape}
\usepackage{natbib}
\usepackage{upgreek}
\usepackage[colorlinks=true,linkcolor=blue, citecolor=blue, filecolor=blue, urlcolor=blue]{hyperref}
\usepackage{listings}

\newcommand{\du}{{\rm d}}
\newcommand{\del}{\vec{\nabla}}

\begin{document}

\title{Self-consistent-field method for triaxial differentiated bodies in hydrostatic equilibrium}

\titlerunning{Self-consistent-field method for triaxial differentiated bodies in hydrostatic equilibrium}

\author{C. Staelen\inst{1} \and J.-M. Hur\'e\inst{2}}
\institute{Universit\'e Paris Cit\'e, CNRS, Institut de Physique du Globe de Paris, F-75005 Paris, France \\ \email{staelen@ipgp.fr} \and Universit\'e de Bordeaux, CNRS, Laboratoire d'Astrophysique de Bordeaux, F-33615 Pessac, France}

\date{Received ??? / Accepted ???}

\abstract{
  Recent observations and models of Haumea and Quaoar suggest that both bodies are triaxial, but their shapes are inconsistent with Jacobi ellipsoids. To determine whether these objects can be at hydrostatic equilibrium, we propose a new numerical code, BALEINES, to study the hydrostatic shape of triaxial differentiated bodies. The fluid mass is assumed to be made of several homogeneous layers, which allowed us to rewrite the gravitational potential as a sum of proper surface integrals. In contrast to the classical self-consistent field method, we did not solve for the mass density, but for the shape of the boundary of all layers, meaning that only one point per layer is needed in the radial direction. The solution is still searched for iteratively. The code was benchmarked against analytical and numerical solutions. As a quick application, we studied the position of the axisymmetric-triaxial bifurcation point of two-layer systems. We show that the deviation from the Meyer bifurcation point in the single-layer case is below $10~\%$ in realistic cases. Based on this result, we conclude that the shape of Quaoar, as obtained in a recent work using a thermophysical model of the surface, is not compatible with a hydrostatic figure of equilibrium. 
}


\maketitle

\section{Introduction}

\begin{figure}[ht]
  \centering
  \includegraphics[width=0.85\linewidth]{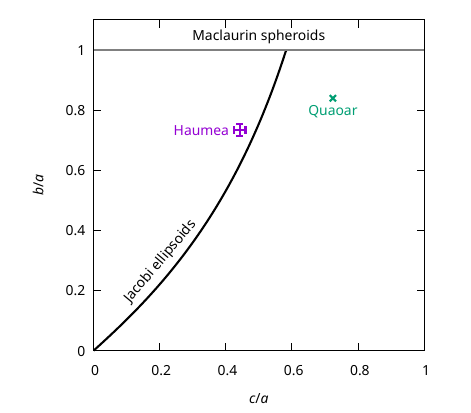}
  \caption{Possible axis ratio couples $(b/a,c/a)$ for ellipsoidal figures of equilibrium, with the Jacobi sequence $(a>b>c)$ as a black line and the Maclaurin sequence $(a=b>c)$ as a gray line. Haumea and Quaoar are placed in the diagram for illustration. The data and error bars for Haumea come from the observations of \citet{ortiz17}, and the data for Quaoar come from the thermophysical model of \citet{kiss2024}.}
  \label{fig:hauqua_vs_jac}
\end{figure}

Following the classical result of \citet{maclaurin42}, who showed that a homogeneous oblate ellipsoid of revolution is a figure of equilibrium, it was long thought that only axisymmetric configurations can describe a self-gravitating rotating fluid mass. Nearly a century later, \citet{jacobi34} proved that homogeneous triaxial ellipsoids may also satisfy hydrostatic equilibrium. Jacobi's theorem establishes that a rigidly rotating homogeneous ellipsoid with semiaxis lengths $a>b>c$ is an equilibrium figure if and only if for given values of $a$ and $b$, the rotation axis coincides with the $c$-axis, whose length has a unique value (see Fig. \ref{fig:hauqua_vs_jac}). \citet{meyer42} further showed that the Maclaurin sequence and the Jacobi sequence share a common member at $c/a\approx 0.58272$. Beyond this Meyer bifurcation point, that is, for $c/a\lesssim 0.58272$, the Maclaurin sequence becomes unstable, and any axisymmetry-breaking perturbation distorts the Maclaurin spheroid into a Jacobi ellipsoid\footnote{Rigourously, this is exact only if the spheroid has a mean to dissipate energy (e.g., viscosity or radiation). Otherwise, the bifurcation occurs even further, for $c/a\approx0.30333$.}. 

For heterogeneous bodies, fundamental results on the existence of barotropic triaxial figures were obtained by \citet{j64} and \citet{im81}. These figures have mostly been studied in the context of galactic dynamics \citep[see, e.g.,][]{vandervoort80} or tidally distorted bodies \citep[e.g.,][]{hachisu86III,et09,whm17}.

More recently, light-curve analyses and stellar-occultation shape measurements of the trans-Neptunian objects Haumea and Quaoar have renewed interest in these heterogeneous equilibrium figures. The shape of Haumea was constrained through stellar occultation by \citet{ortiz17}, who determined that the best-fitting ellipsoid had ${a=1161\pm30~\mathrm{km}}$, ${b=852\pm4~\mathrm{km}}$, and ${c=513\pm16~\mathrm{km}}$. Using another equation set that took photometry into account, \citet{kondra18} obtained the values ${a=1082\pm15~{\rm km}}$, ${b=836\pm5~\mathrm{km}}$, and ${c=511\pm13~\mathrm{km}}$. For Quaoar, \citet{kiss2024} have recently developed a thermophysical model of its surface and found that the best match to the observed thermal light curve corresponds to a triaxial figure with $a=1.29\times10^3~{\rm km}$, $b=1.08\times10^3~{\rm km}$, and $c=9.32\times10^2~{\rm km}$. However, as illustrated in Fig. \ref{fig:hauqua_vs_jac}, the resulting axis ratios $b/a$ and $c/a$ for both bodies are inconsistent with Jacobi ellipsoids. To explain this discrepancy in the case of Haumea, \citet{ortiz17} proposed two possibilities: either Haumea is not in hydrostatic equilibrium, meaning rigid-body forces are required to properly model its shape, or it is not homogeneous, but instead composed of multiple layers. The same alternatives might be considered for Quaoar. 

The likelihood that Haumea is hydrostatic was studied by \citet{ddp19} and \cite{noviello22} in the case of a two-layer planet with a silicate core embedded in an icy mantle. These authors explored how the internal differentiation enlarges the range of possible equilibrium shapes. Using the self-consistent-field (SCF) method, in particular, the three-dimensional algorithm of \citet{hachisu86III}, they obtained configurations that matched the light curve derived by \citet{lbs14}, but remained inconsistent with, but closer to, the shape inferred by \citet{ortiz17}. Consequently, it is still unclear whether a differentiated hydrostatic model can fully reproduce the current observations of Haumea, or if a hydrostatic configuration is likely for Quaoar. If models like this were to be found, they would impose strong constraints on the layer compositions and relative sizes.

However, the algorithm of \citet{hachisu86III} uses a regular grid in spherical coordinates, which requires a high radial resolution to precisely determine the shape of the layer boundaries. This constraint leads to high computational costs and can be an obstacle to a detailed exploration of the parameter space. We aim to introduce a numerical method designed to overcome this limitation and to investigate the equilibrium figures of triaxial differentiated bodies composed of two or more homogeneous layers. In particular, we fully exploit the hypothesis of a piecewise homogeneous system to optimize the number of radial points needed (i.e., one per layer), and we thus greatly reduce the computation time compared to classical SCF methods.

The paper is organized as follows: In Sect. \ref{sec:theory} we detail the assumptions and the equation set. In Section \ref{sec:scf} we adapt the SCF method to the problem of differentiated bodies. The code BALEINES is presented in Sect. \ref{sec:tests}, and we report successful comparisons with analytical solutions from the Maclaurin and Jacobi sequences and numerical solutions from \citet{hachisu86III}, \citet{descamps15}, \citet{ddp19}, and \citet{noviello22}. In Sect. \ref{sec:examples} we discuss equilibrium sequences of two-layer bodies and the effect of differentiation on the axisymmetric-triaxial bifurcation, with an application to the shape of Quaoar. Our concluding remarks and some perspectives are given the last section.

\section{Theoretical background}\label{sec:theory}

\subsection{Classical SCF method}

The SCF method originated in quantum chemistry with the Hartree-Fock method, which provides an approximation for the wave function and the fundamental level energy of multi-atom systems. It was then adapted to the context of a self-gravitating rotating fluid by \citet{om68}. Major improvements were brought by \citet{hachisu86III} to better deal with very fast rotators.

The equilibrium of a rigidly rotating and self-gravitating fluid is given by a Bernoulli-like equation,
\begin{equation}\label{eq:bernoulli}
  \int\frac{\du P}{\rho} - \frac12(\vec{\varOmega}\times\vec{r})^2 + \varPsi(\vec{r}) = C,
\end{equation}
where $P$ is the thermodynamic pressure, $\rho$ is the mass density, $\vec{\varOmega}$ is the rotation rate vector, $\vec{r}$ is the position vector, and $C$ is a constant to be determined, usually from a point on the free surface of the fluid. The gravitational potential, $\varPsi$, is given by 
\begin{equation}\label{eq:psi_newton}
  \varPsi(\vec{r}) = -G\iiint_{\cal V}\du^3\vec{r'} \, \frac{\rho(\vec{r'})}{|\vec{r}-\vec{r'}|},
\end{equation}
where $\cal V$ is the volume of the fluid, and $G$ is the gravitational constant. 

Thus, as remarked by \citet{od03}, Eq. \eqref{eq:bernoulli} is in fact a nonlinear Hammerstein integral equation with unknown $\rho$; this is then a fixed-point problem. When the body greatly deviates from a sphere, this problem must be solved with numerical iterative methods. After the choice of an initial guess, the SCF method can be summarized as a cycle with two main steps. First, we compute $\varPsi$ to obtain a new guess on the mass density through Eq. \eqref{eq:bernoulli}, and second, we use this new guess on $\rho$ to compute a new gravitational potential. The procedure is repeated until the solution has converged. The main difference in the implementation of this method is the computation of the gravitational potential, which is a great problem because of the point-mass divergence in the integrand (see Eq. \eqref{eq:psi_newton}). This problem is usually avoided either by using the multipolar expansion of the potential \citep{om68,hachisu86III,kny10} or by solving the Poisson equation $\nabla^2\varPsi=4\uppi G\rho$ \citep{ka16,hh17}.

\subsection{The gravitational potential of homogeneous bodies}\label{ssec:gravpot}

In homogeneous bodies, the mass density is a constant, and it might seem as if the SCF method became trivial. This is clearly not so simple because the volume of the fluid, particularly the shape of the free surface, is unknown and the classical SCF method still uses the same cycle. The main difficulty in this case is the detection of the free surface, which becomes a key point for the numerical precision. When classical grids are used, the free surface can be described precisely using a coded Freeman chain \citep{hh17}. Another approach, introduced by \citet{ansorg03} in the context of axisymmetric bodies, is to use the assumption of homogeneity to use the Kovalevskaya line integral for the potential \citep{kowalewsky85}. As a consequence, the method does not solve for the mass density at each node of a grid, which would give the shape of the free surface, but for the shape itself. This framework then makes the problem one-dimensional and greatly reduces the computation time with respect to classical two-dimensional axisymmetric approaches. When the body has no axial symmetry, a similar transformation can be made. When we assume that $\rho$ is constant inside the fluid, the volume integral in Eq. \eqref{eq:psi_newton} can indeed be reduced to a surface integral \citep{gausswerke5,tisserand91},
\begin{equation}\label{eq:psi_vec}
  \varPsi(\vec{r}) = \frac12G\rho\oiint_{{\cal S}}\du^2\vec{r'} \, \vec{n}\cdot\frac{\vec{r}-\vec{r'}}{|\vec{r}-\vec{r'}|},
\end{equation}
where ${\cal S}$ is the closed surface bounding $\cal V$, and $\vec{n}$ is the outward normal to ${\cal S}$ at $\vec{r'}$. This result can be proved with the divergence theorem. A first consequence of this expression is that the kernel remains finite at coincidence, that is, for $\vec{r}=\vec{r'}$. Indeed, $\vec{r'}$ is restricted to the surface, and as $\vec{r'}\rightarrow\vec{r}$, $\vec{r}-\vec{r'}$ thus tends to a vector tangent to the surface at $\vec{r}$, which is clearly orthogonal to the normal $\vec{n}$; the kernel is thus null at coincidence. As a consequence, the point-mass divergence problem is avoided, and direct integration can be made if $\vec{r}$ belongs to the surface. 

\subsection{Hydrostatic equilibrium of piecewise homogeneous systems}

We considered a differentiated body made of $L$ homogeneous layers. We labeled each layer with an index $\ell$, where $\ell=1$ is the innermost layer and $\ell={L}$ is the outermost one. Let ${\cal S}_{L}$ be the free surface of the fluid and ${\cal S}_{\ell}$ be the interface between layers $\ell$ and $\ell+1$ ($1\leq\ell\leq L-1$). As $\varPsi$ is linear with the mass density, the potential of a piecewise homogeneous mass distribution can be deduced from the superposition principle, namely
\begin{equation}
  \varPsi(\vec{r}) = \frac12G\sum_{\ell=1}^{L}(\rho_\ell-\rho_{\ell+1})\oiint_{{\cal S}_{\ell}}\du^2\vec{r'} \, \vec{n}\cdot\frac{\vec{r}-\vec{r'}}{|\vec{r}-\vec{r'}|},
\end{equation}
where $\rho_{\ell}$ is the mass density of layer $\ell$. We set $\rho_{{L}+1}=0$ for convenience. 

We assumed that the polar axis coincides with the Cartesian $({\rm O}z)$ axis, $\rm O$ being the barycenter of the fluid, while the equatorial major and minor axes coincide with $({\rm O}x)$ and $({\rm O}y)$, respectively. We worked with spherical coordinates $(r,\theta,\varphi)$, where $r$ is the distance to the center of mass of the body, $\theta$ is the colatitude, and $\varphi$ is the azimuthal angle. These coordinates are suitable for almost ellipsoidal figures, that is, for figures in which any ray drawn from the center of mass intersects the boundary exactly once. This is clearly not the case for toroids\footnote{It can be argued that cylindrical coordinates $(\varpi,\varphi,z)$ are a better choice, as toroids can easily be described in this system; this choice was made by \citet{ansorg03}. However, if $z=\zeta(\varpi,\varphi)$ is the equation of the surface, then it would be required to compute $\partial\zeta/\partial\varpi$ at each point for the potential. At the equator, this derivative becomes infinite, which is a problem for the numerical treatment. This is not the case with spherical coordinates, where $\partial s/\partial \theta$ is always finite.}. With this assumption, we represented the boundary $\cal S_{\ell}$ by a function $s_{\ell}(\theta,\varphi)$, so that the equation of $\cal S_{\ell}$ is $r=s_\ell(\theta,\varphi)$. The potential reads 
\begin{align}
  &\varPsi(r,\theta,\varphi) = \\
  &\mkern+50mu\frac12G\sum_{\ell=1}^{L}(\rho_\ell-\rho_{\ell+1})\int_0^{\uppi}\du\theta'\int_0^{2\uppi} \du\varphi' \frac{\nu(\theta',\varphi';s_{\ell};r,\theta,\varphi)}{\sigma(\theta',\varphi';s_{\ell};r,\theta,\varphi)},\notag
\end{align}
where
\begin{align}\label{eq:psi_nu}
  \nu&(\theta',\varphi';s_{\ell};r,\theta,\varphi) = \\
  &rs_{\ell}^2(\theta',\varphi')\sin(\theta')\big[\sin(\theta)\sin(\theta')\cos(\varphi-\varphi')+\cos(\theta)\cos(\theta')\big]\notag\\
  &-rs_{\ell}(\theta',\varphi')\frac{\partial s_{\ell}}{\partial \theta'}\sin(\theta')\big[\sin(\theta)\cos(\theta')\cos(\varphi-\varphi')\notag\\
  &\mkern+300mu-\cos(\theta)\sin(\theta')\big]\notag\\
  &-rs_{\ell}(\theta',\varphi')\frac{\partial s_{\ell}}{\partial \varphi'}\sin(\theta)\sin(\varphi-\varphi') - s_{\ell}^3(\theta',\varphi')\sin(\theta')\notag
\end{align}
and
\begin{align}
  \sigma^2&(\theta',\varphi';s_{\ell};r,\theta,\varphi) = r^2+s_{\ell}^2(\theta',\varphi')\notag\\
  &-2rs_{\ell}(\theta',\varphi')\big[\sin(\theta)\sin(\theta')\cos(\varphi-\varphi')+\cos(\theta)\cos(\theta')\big].
\end{align}

The hydrostatic equilibrium is governed by $L$ Bernoulli-like equation, one per layer. Inside layer $\ell$, Eq. \eqref{eq:bernoulli} reads
\begin{equation}
  \frac{P(\vec{r})}{\rho_{\ell}} - \frac12(\vec{\varOmega}\times\vec{r})^2 + \varPsi(\vec{r}) = C_{\ell},
\end{equation}
where the integral over the pressure has been readily computed, and $C_{\ell}$ is a constant.

In absence of an ambient medium, the pressure vanishes along the free surface ${\cal S}_{L}$, namely
\begin{equation}\label{eq:bernoulli_free}
  - \frac{\varOmega^2s_{L}^2(\theta,\varphi)\sin^2(\theta)}{2} + \varPsi\big(s_{L}(\theta,\varphi),\theta,\varphi\big) = C_{L},
\end{equation}
with $\varOmega = |\vec{\varOmega}|$ being the rotation rate. Moreover, the pressure is continuous at the interface ${\cal S}_{\ell}$, which leads to
\begin{equation}\label{eq:bernoulli_inter}
  - \frac{\varOmega^2s_{\ell}^2(\theta,\varphi)\sin^2(\theta)}{2} + \varPsi\big(s_{\ell}(\theta,\varphi),\theta,\varphi\big) = \frac{\rho_{\ell}C_{\ell}-\rho_{\ell+1}C_{\ell+1}}{\rho_{\ell}-\rho_{\ell+1}}.
\end{equation}
Equations \eqref{eq:bernoulli_free} and \eqref{eq:bernoulli_inter} merely state that ${\cal S}_{\ell}$ are level surfaces, that is, surfaces at which the pressure and the sum of the centrifugal and gravitational potential are constant. 
 
As $\varPsi$ depends only on the shape of the surfaces, the problem of the hydrostatic equilibrium of a differentiated body then consists of finding the set of functions $\{{s}_{\ell}(\theta,\varphi)\}$ such that Eqs. \eqref{eq:bernoulli_free} and \eqref{eq:bernoulli_inter} are satisfied simultaneously for all $\ell$; this formulation of the problem clearly recalls the theory of equilibrium figures \citep[e.g.,][]{zt78,lan82}. Clearly, Eqs. \eqref{eq:bernoulli_free} and \eqref{eq:bernoulli_inter} are nonlinear coupled integral equations for the set of functions $\{{s}_{\ell}\}$ and form a fixed-point problem. As for the classical SCF method, iterative methods are required to treat the problem in the general case.

\section{The self-consistent-field method for differentiated bodies}\label{sec:scf}

As only the shapes of surfaces are needed at a given step, only $L$ points are needed in the radial direction to describe the system for a given numerical resolution in $(\theta,\varphi)$. For a small number of layers, this sort of adaptive meshing reduces the computation time with respect to a classical SCF method, which needs a great number of points to precisely determine the interfaces between layers. This advantage in time is clearly reduced as the number of layers increases, so that the method probably has similar performances as the classical SCF method for a continuously heterogeneous system (i.e., ${L} \gg 1$)\footnote{Note that using this method directly to describe continuous systems is equivalent to approximating an integral with its Riemann sum, whose accuracy is only of the order of $1/{L}$ -- this is however done in the Concentric Maclaurin Spheroids method \citep{hub13}; see also \citet{dc18}. For a better precision in this case, one may want to replace $\sum_{\ell}(\hat\rho_\ell-\hat\rho_{\ell+1})$ in Eq. \eqref{eq:psi_dedi} by a trapezoidal rule or any other numerical integration method.}. 

\subsection{The dimensionless problem}

For the numerical treatment, it is more suitable to work with dimensionless quantities. We define
\begin{equation}
  \left\{
    \begin{aligned}
      &\hat{r} = r/a_L,\\
      &\hat{s}_{\ell} = s_{\ell}/a_L,\\
      &\hat\rho_\ell = \rho_\ell/\rho_{L},\\
      &\hat\varPsi = \varPsi/(G\rho_{L}a_L^2),\\
      &\hat\varOmega^2 = \varOmega^2/(G\rho_{L}),\\
      &\hat{C}_\ell = C_\ell/(G\rho_{L}a_L^2),
    \end{aligned}
  \right.
\end{equation}
where $a_L = s_{L}(\uppi/2,0)$ is the equatorial major semiaxis of the whole mass. In dimensionless spherical coordinates, the gravitational potential reads
\begin{equation}\label{eq:psi_dedi}
  \hat\varPsi(\hat{r},\theta,\varphi) = \frac12\sum_{\ell=1}^{L}(\hat\rho_\ell-\hat\rho_{\ell+1})\int_0^{\uppi}\du\theta'\int_0^{2\uppi}\du\varphi' \kappa_\ell(\theta',\varphi';\hat{r},\theta,\varphi), 
\end{equation}
where the kernel $\kappa_\ell$ is 
\begin{align}\label{eq:kernel}
  \kappa_\ell(\theta',\varphi';\hat{r}&,\theta,\varphi)=\frac{\nu(\theta',\varphi';s_{\ell};r,\theta,\varphi)}{\sigma(\theta',\varphi';s_{\ell};r,\theta,\varphi)}\times\frac{1}{a_L^2}.
\end{align}

For convenience, let 
\begin{equation}
  \hat\varPsi_\ell(\theta,\varphi) = \hat\varPsi\big(s_{\ell}(\theta,\varphi),\theta,\varphi\big)
\end{equation}
be the gravitational potential along ${\cal S}_{\ell}$. Equations \eqref{eq:bernoulli_free} and \eqref{eq:bernoulli_inter} can be rewritten as 
\begin{equation}\label{eq:bernoulli_dedi}
  \hat\varPsi_\ell(\theta,\varphi)-\frac12\hat\varOmega^2\hat{s}_{\ell}^2(\theta,\varphi)\sin^2(\theta) - \frac{\hat\rho_{\ell}\hat{C}_{\ell}-\hat\rho_{\ell+1}\hat{C}_{\ell+1}}{\hat\rho_{\ell}-\hat\rho_{\ell+1}} = 0,
\end{equation}
$\forall \ell \in \llbracket 1,{L}\rrbracket$, where we set $\hat{C}_{L+1}=0$ for convenience.

\subsection{Description of the cycle}

For this SCF method, we started with a set of initial functions $\{\hat{s}_{\ell}^{(0)}(\theta,\varphi)\}$. We chose to start with a set of ellipsoidal surfaces with dimensionless semiaxis length $\hat{a}_\ell \geq \hat{b}_\ell \geq \hat{c}_\ell$, that is,
\begin{equation}
  \hat{s}_{\ell}^{(0)}(\theta,\varphi) = \frac{1}{\sqrt{\dfrac{\sin^2(\theta)\cos^2(\varphi)}{\hat{a}_{\ell}^2}+\dfrac{\sin^2(\theta)\sin^2(\varphi)}{\hat{b}_{\ell}^2}+\dfrac{\cos^2(\theta)}{\hat{c}_{\ell}^2}}}.
\end{equation}
These initial ellipsoids were taken to be similar, that is, $\hat{b}_{\ell}/\hat{a}_{\ell}$ and $\hat{c}_{\ell}/\hat{a}_{\ell}$ were the same for each starting surface. The mass density ratios from one layer to the layer above, $\hat\rho_{\ell}/\hat\rho_{\ell+1},\, \ell\in\llbracket 1,L-1\rrbracket$, were set as inputs. The other inputs are detailed in Sect. \ref{ssec:om2cte}.

Then, at iteration $t>0$ of the cycle, we computed
\begin{enumerate}[(i)]
  \item the gravitational potential along each interface, $\hat\varPsi_{\ell}^{(t)}$, using Eq. \eqref{eq:psi_dedi};
  \item the constants $\hat{C}^{(t)}_{\ell}$ and the rotation rate squared, $\hat\varOmega^{2(t)}$, that is, the method detailed in Sect. \ref{ssec:om2cte}; and
  \item a new set of surfaces $\{\hat{s}_{\ell}^{(t)}(\theta,\varphi)\}$ were obtained by searching for the zero of Eq. \eqref{eq:bernoulli_dedi} for all $(\theta,\varphi)$ of the grid.
\end{enumerate}
These steps were repeated until convergence, that is, when all quantities no longer varied from one step to the next within a given threshold, $\epsilon_0$. Thus, as a criterion, we used 
\begin{equation}\label{eq:threshold}
  \epsilon = \max\left|\hat{s}_{\ell}^{(t+1)}(\theta,\varphi)-\hat{s}_{\ell}^{(t)}(\theta,\varphi)\right| < \epsilon_0.
\end{equation}
We set $\epsilon_0$ to $10^{-14}$, that is, about $100$ times the machine epsilon in double precision. This threshold can clearly be set higher (e.g., $\epsilon_0=10^{-8}$ or $\epsilon_0=10^{-6}$) to reduce the computation time; the value of $10^{-14}$ merely ensures that the solution is numerically stable.

\subsection{Inputs and computation of the rotation rate and the constants}\label{ssec:om2cte}

As mentioned in the previous paragraph, computing the constants $\hat{C}_{\ell}$ and the rotation rate squared $\varOmega^2$ is required at each iteration. We first considered a homogeneous body to facilitate the discussion. We then determined a single constant, $\hat{C}_1$, and the square rotation rate. Following \citet{hachisu86III}, classical SCF methods hold two points of the free surface fixed by requiring that the Bernoulli-like equation is automatically satisfied at these points, which yields the values of $\hat{C}_1$ and $\hat\varOmega^2$. When $L>1$, $\hat{C}_L$ and $\hat\varOmega^2$ are still determined by holding two points of the free surface, ${\cal S}_L$, fixed as previously. For a deeper layer $\ell$, only $\hat{C}_{\ell}$ has to be determined. Thus, one point of the surface ${\cal S}_{\ell}$ needs to be held fixed. As a consequence, in addition to the mass density ratios $\hat\rho_{\ell}/\hat{\rho}_{\ell+1}$, the position of $L+1$ points, $2$ on the free surface and $1$ on the other boundaries, are needed to solve the problem. 

Alternatively, we might impose the value of the geodetic parameter, ${\tt m} = \varOmega^2R^3/(GM)$, $R$ being the volumetric radius and $M$ the mass, to determine the value of $\hat\varOmega^2$ at each iteration. Therefore, only one point had to be fixed on the free surface in this formulation. We might also wish to impose the volume fraction of each layer to track the evolution of a system with time. As a consequence, no point would be fixed on the inner boundaries because it would be rescaled from one iteration to the next. 

The choice of inputs then depends on the physical problem that is to be solved. The implementation of all these formulations is detailed in the following paragraphs. 

\subsubsection{Imposing the polar axis lengths}

In classical SCF methods, the usual choice is to hold the polar semiaxis length of each layer and the equatorial major semiaxis length of the free surface fixed. The latter is implicitly assumed by our choice of normalization. To do this, the rotation rate and the constants are given by
\begin{equation}
  \hat\varOmega^2 = \hat\varPsi_L(\uppi/2,0)-\hat\varPsi_L(0,0),
\end{equation}
where we used $\hat{s}_L^2(\uppi/2,0)=1$, and
\begin{equation}
  \left\{
    \begin{aligned}
      &\hat{C}_L = \hat\varPsi_L(0,0)\\
      &\frac{\hat\rho_{\ell}\hat{C}_{\ell}-\hat\rho_{\ell+1}\hat{C}_{\ell+1}}{\hat\rho_{\ell}-\hat\rho_{\ell+1}} = \hat\varPsi_{\ell}(0,0).
    \end{aligned}
  \right.
\end{equation}
These expressions ensure that the Bernoulli-like equation is always satisfied at the points we wished to hold fixed. The inputs of this formulation are thus the polar semiaxis length of each layer, $\hat{c}_{\ell}$, and the mass density ratios $\rho_{\ell}/\rho_{\ell+1}$.

\subsubsection{Imposing the equatorial minor axis lengths}

Another possible choice is to hold the equatorial minor semiaxis of each layer and the equatorial major semiaxis length of the free surface fixed. This can be useful when the equatorial minor axis is better constrained that the polar axis, or when the bifurcation from an axisymmetric figure to a triaxial figure is to be studied, which is easier when the equatorial minor semiaxis of the surface is an input. This is described in Sect. \ref{ssec:bifurcation} for two-layer systems. In this case, the expressions are very similar to the previous case. We have
\begin{equation}
  \hat\varOmega^2 = \frac{\hat\varPsi_L(\uppi/2,0)-\hat\varPsi_L(\uppi/2,\uppi/2)}{1-\hat{s}_{L}^2(\uppi/2,\uppi/2)},
\end{equation}
where
\begin{equation}
  \left\{
    \begin{aligned}
      &\hat{C}_L = \hat\varPsi_L(\uppi/2,\uppi/2) - \frac12\hat\varOmega^2 \hat{s}_L^2(\uppi/2,\uppi/2),\\
      &\frac{\hat\rho_{\ell}\hat{C}_{\ell}-\hat\rho_{\ell+1}\hat{C}_{\ell+1}}{\hat\rho_{\ell}-\hat\rho_{\ell+1}} = \hat\varPsi_{\ell}(\uppi/2,\uppi/2) - \frac12\hat\varOmega^2 \hat{s}_{\ell}^2(\uppi/2,\uppi/2).
    \end{aligned}
  \right.
\end{equation}
The inputs of this formulation are the equatorial minor semiaxis length of each layer, $\hat{b}_{\ell}$ and the mass density ratios $\rho_{\ell}/\rho_{\ell+1}$.

\subsubsection{Geodetic parameter}

With the dimensionless quantities we used, the geodetic parameter, $\tt m$, reads
\begin{equation}
  {\tt m} = \frac{\hat\varOmega^2\hat{R}^3}{\hat{M}},
\end{equation}
with $\hat{R} = R/a_L$. This value is often used in theory of figures as an input. To do this, the Bernoulli constants must be written
\begin{equation}
  \left\{
    \begin{aligned}
      &\hat{C}_{L} = \hat\varPsi_{L}(\uppi/2,0) - \frac12{\tt m}\frac{\hat{M}}{\hat{R}^3} \hat{s}_{L}^2(\uppi/2,0),\\
      &\frac{\hat\rho_{\ell}\hat{C}_{\ell}-\hat\rho_{\ell+1}\hat{C}_{\ell+1}}{\hat\rho_{\ell}-\hat\rho_{\ell+1}} = \hat\varPsi_{\ell}(\uppi/2,0) - \frac12{\tt m}\frac{\hat{M}}{\hat{R}^3} \hat{s}_{\ell}^2(\uppi/2,0).
    \end{aligned}
  \right.
\end{equation}
These expressions also hold the points at the end of the equatorial major semiaxis of each layer fixed. As the length of the polar and equatorial minor semiaxes of the free surface may vary from an iteration to another, holding the length of the equatorial major semiaxes fixed probably is the safest choice.

These equations then require the computation of the dimensionless mass, $\hat{M}$, and the dimensionless volumetric radius, $\hat{R}$, at each iteration of the cycle. The inputs of this formulation are the equatorial major semiaxis length of each layer, $\hat{a}_{\ell}$, the geodetic parameter, $\tt m$, and the mass density ratios $\rho_{\ell}/\rho_{\ell+1}$.

\subsubsection{Imposing the volume fraction}

The volume fraction of the layers, $\nu_{\ell}$, can also be imposed to be constant from one iteration to the next. To do this, we first determined the surfaces at which pressure is continuous at each iteration, $t$, of the cycle. However, there is no reason that the volume of the layer $\ell$ determined in this way corresponds to a fraction $\nu_{\ell}$ of the total volume and has to be rescaled accordingly. Let $\hat{s}_{\ell}^{(t+1/2)}$ be the surface equation of layer $\ell$, where $t+1/2$ is an intermediate step between $t$ and $t+1$, corresponding to the rescaling procedure. As the dimensionless equatorial major semiaxis length of the free surface is always set to 1, we directly have $\hat{s}_L^{(t+1)} = \hat{s}_L^{(t+1/2)}$, and the total volume reads
\begin{align}
  \hat{V}^{(t+1)} = \frac13\int_0^{\uppi}\du\theta\sin(\theta)\int_0^{2\uppi}\du\varphi \left[\hat{s}_{L}^{(t+1)}(\theta,\varphi)\right]^3.
\end{align}
Similarly, the volume bounded by the surface ${\cal S}_{\ell}$ at the intermediate step $t+1/2$ is given by
\begin{align}
  \hat{V}_{\ell}^{(t+1/2)} = \frac13\int_0^{\uppi}\du\theta\sin(\theta)\int_0^{2\uppi}\du\varphi \left[\hat{s}_{\ell}^{(t+1/2)}(\theta,\varphi)\right]^3.
\end{align}
As a consequence, the rescaled surface equation of layer $\ell$ can be written
\begin{equation}\label{eq:cvf}
  \hat{s}_{\ell}^{(t+1)}(\theta,\varphi) = \hat{s}_{\ell}^{(t+1/2)}(\theta,\varphi)\left[\left(1-\sum_{j=\ell+1}^{L}\nu_j\right)\frac{\hat{V}^{(t+1)}}{\hat{V}_j^{(t+1/2)}}\right]^{1/3}.
\end{equation}

This formulation needs to be completed with one of the previous formulations to compute the rotation rate and the constants; in addition to its inputs, imposing the volume fraction requires the value of the volume fraction of each layer, $\nu_{\ell}$. This allowed us to study sequences of equilibrium in which the layers have the same volume (or equivalently, the same mass), which could model the quasi-static evolution of a hydrostatic body with a varying rotation rate. Some examples of such sequences are given in Sect. \ref{ssec:sequence}.

\subsection{Numerical computation of the gravitational potential}\label{ssec:psinum}

As mentioned in the previous section, the main advantage of the surface integral in Eq. \eqref{eq:psi_vec} is that the integrand is fully finite and allows us to compute the gravitational potential on the surface directly. However, this advantage comes with two problems.

First, the integrand is continuous and finite on the surface, but its derivatives are discontinuous at $\vec{r}=\vec{r}'$, with the consequence that the numerical precision of standard high-accuracy integration methods (Gauss-Legendre, Clenshaw-Curtis, etc.) designed for continuous and infinitely derivable functions is limited. The trapezoidal rule is not affected by this problem, but can only achieve a precision of $\sim1/N^2$, where $N$ is the number of nodes used for the integration. 

\begin{figure}[ht]
  \centering
  \includegraphics[width=0.85\linewidth]{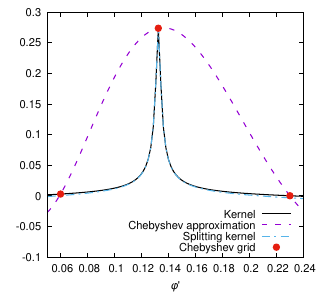}
  \caption{Kernel function given by Eq. \eqref{eq:kernel} for an ellipsoid with $a=1$, $b=0.9$, $c=0.6$ at $(\theta,\varphi)\approx(0.632,0.132)$, corresponding to a node of the Chebyshev grid, and for $r=1.001\, s(\theta,\varphi)$. The kernel, $\kappa_{\ell}$, is plotted as a function of $\varphi'$ for $\theta'=\theta$ and zoomed-in on the lobe. The Chebyshev grid has $N+1=17$ nodes. The dashed line corresponds to the Chebyshev approximation of the kernel, and the dashed-and-dotted line shows the kernel of the concentric and coincident sphere, corresponding to $\kappa_{\ell,0}$ in Eq. \eqref{eq:splitting}.}
  \label{fig:lobe}
\end{figure}

Second, it can be proved that when the potential is not computed exactly on the surface, but slightly above or below it, the kernel $\kappa$ exhibits a lobe at $(\theta',\varphi')=(\theta,\varphi)$ whose width decreases with $|\vec{r}-\vec{r'}|$. Thus, for points that are very close to the surface, this lobe is poorly resolved on the numerical grid, and its contribution is slightly overestimated. This produces a discontinuity of the potential at the surface, which clearly is an artifact produced by the poorly accounted for lobe. 

A possible method for solving both problems at once is to split the kernel in two, to isolate the problematic part of the integrand, and integrate it analytically \citep{hure05}. When $\kappa_{\ell,0}$ is the splitting kernel, the method simply reads 
\begin{align}\label{eq:splitting}
  \oiint_{{\cal S}_{\ell}} \du^2\vec{r'} \kappa_{\ell}(\vec{r},\vec{r'}) = 
  \oiint_{{\cal S}_{\ell}} \du^2\vec{r'} &\left[\kappa_{\ell}(\vec{r},\vec{r'})-\kappa_{\ell,0}(\vec{r},\vec{r'})\right]\\ 
  &\mkern+90mu+\oiint_{{\cal S}_{\ell}} \du^2\vec{r'} \kappa_{\ell,0}(\vec{r},\vec{r'}),\notag
\end{align}
where the last term is expected to be analytically integrable at least once. To address the problem at coincidence at least partly, it can be proved that a the splitting kernel has to share its value and its first two derivatives at $(\theta,\varphi)$ with the original kernel. However, this makes the mathematical treatment more difficult, and most of the solutions we found for the moment were not general enough to cover a satisfying number of cases. For now, we chose to focus on the lobe, which is more problematic for the SCF-cycle. To keep a simple method that works well enough in all cases, the splitting kernel in this work was the kernel associated with a sphere, concentric and coincident with the fluid at $(\theta,\varphi)$, that is,
\begin{align}
  &\kappa_{\ell,0}(\hat{r},\theta,\varphi;\theta',\varphi') = \\
  &\mkern+20mu\frac{\hat{r}\hat{R}_{\ell}^2\sin(\theta')[\sin(\theta)\sin(\theta')\cos(\varphi-\varphi')+\cos(\theta)\cos(\theta')]}{\sqrt{\hat{r}^2+\hat{R}_{\ell}^2-2\hat{r}\hat{R}_{\ell}[\sin(\theta)\sin(\theta')\cos(\varphi-\varphi')+\cos(\theta)\cos(\theta')]}},\notag
\end{align}
where $\hat{R}_{\ell}=\hat{s}_{\ell}(\theta,\varphi)$ is the radius of the aforementioned sphere. Thus, the surface integral of $\kappa_{\ell,0}$ in Eq. \eqref{eq:splitting} corresponds to the gravitational potential of a sphere, namely
\begin{equation}
  \oiint_{{\cal S}_{\ell}} \du^2\vec{r'} \kappa_{\ell,0}(\vec{r},\vec{r'}) = -\frac23\uppi\times\left\{
     \begin{aligned}
      &(3\hat{R}_{\ell}^2-\hat{r}^2),&\hat{r}\leq\hat{R}_{\ell},\\
      &2\frac{\hat{R}_{\ell}^3}{\hat{r}},&\hat{r}>\hat{R}_{\ell}.
     \end{aligned}
  \right.
\end{equation}
This splitting does not enhance the accuracy of the quadratures when the field point is on the surface, but, as shown in Fig. \ref{fig:lobe}, it allowed us to take the lobe better into account when the field point was slightly above or below the surface.

\subsection{Computation of the outputs}

After convergence was reached, the outputs of the cycle itself were the dimensionless rotation rate, $\hat\varOmega$, the shape of the surfaces, $\hat{s}_{\ell}(\theta,\varphi)$, and the Bernoulli constants in each layer, $\hat{C}_{\ell}$. They can be used to compute several properties of the solution, such as the mass, $M$, the moment of inertia with respect to the rotation axis, $I$, the angular momentum, $J$, or the pressure at all interfaces and at the center, $P_{\ell}$ and $P_{\rm c}$ respectively. The dimensionless mass, $\hat{M}=M/(\rho_La_L^3)$, moment of inertia, $\hat{I}=I/(\rho_La_L^5)$, and angular momentum, $J/[G\rho_L^3a_L^{10}]^{1/2}$, are given by
\begin{equation}
  \left\{
    \begin{aligned}
      \hat{M} &= \frac13\sum_{\ell=1}^{L}(\hat\rho_{\ell}-\hat\rho_{\ell+1})\int_0^{\uppi}\du\theta\sin(\theta)\int_0^{2\uppi}\du\varphi \hat{s}_{\ell}^3(\theta,\varphi),\\
      \hat{I} &= \frac15\sum_{\ell=1}^{L}(\hat\rho_{\ell}-\hat\rho_{\ell+1})\int_0^{\uppi}\du\theta\sin^3(\theta)\int_0^{2\uppi}\du\varphi \hat{s}_{\ell}^5(\theta,\varphi),\\
      \hat{J} &= \hat{I} \hat\varOmega.
    \end{aligned}
  \right.
\end{equation}
The pressure along surface ${\cal S}_{\ell}$ is given by
\begin{align}
  \hat{P}_{\ell} = \frac{P_{\ell}}{G\rho_L^2a_L^2} &= \hat{\rho}_{\ell}\left[\hat{C}_{\ell} - \hat{\varPsi}_{\ell}(0,0)\right] \notag \\
  &= \hat{\rho}_{\ell}\hat{\rho}_{\ell+1}\frac{\hat{C}_{\ell}-\hat{C}_{\ell+1}}{\hat{\rho}_{\ell}-\hat{\rho}_{\ell+1}}, \qquad \hat\rho_{\ell} \neq \hat\rho_{\ell+1}
\end{align}
and the pressure at the center by
\begin{equation}
  \hat{P}_{\rm c} = \frac{P_{\rm c}}{G\rho_L^2a_L^2} = \hat{\rho}_1 \left[\hat{C}_1 - \hat{\varPsi}(0,0,0)\right].
\end{equation}

The ratio of kinetic to gravitational energies, $|T/W|$, often called the stability indicator in the context of SCF methods, can be computed as the ratio of the dimensionless kinetic energy, $\hat{T}=T/(G\rho_L^2a_L^5)$, 
\begin{equation}
  \hat{T} = \frac{\hat\varOmega^2}{10}\sum_{\ell=1}^{L} (\rho_\ell-\rho_{\ell+1})\int_0^{\uppi}\du\theta\,\sin^3(\theta)\int_0^{2\uppi}\du\varphi \, \hat{s}_{\ell}^5(\theta,\varphi),
\end{equation}
and the dimensionless gravitational energy, $\hat{W}=W/(G\rho_L^2a_L^5)$, 
\begin{equation}
  \hat{W} = \sum_{\ell=1}^{L} (\hat\rho_\ell-\hat\rho_{\ell+1})\int_0^{\uppi}\du\theta\,\sin(\theta)\int_0^{2\uppi}\du\varphi\,\frac{\hat{s}_{\ell}^3(\theta,\varphi)}{5}\hat\varPsi_{\ell}(\theta,\varphi),
\end{equation}
The proof of this last expression, which we did not find in the literature, is given in Appendix \ref{app:w_proof}.

\section{Numerical tests}\label{sec:tests}

Using the equations and the cycle described in the previous sections, we built a Fortran90 code, BALEINES (for B{\footnotesize ALEINES} is an algorithm looking for equilibria iteratively for nested ellipsoids using surface integrals), to solve the shape of the layers of a triaxial differentiated fluid mass at hydrostatic equilibrium. In this code, the equations of the surfaces, $\hat{s}_{\ell}$, are expanded on Chebyshev polynomials and the surface integrals are evaluated with Clenshaw-Curtis quadratures. The two methods share the same numerical resolution, $N$, corresponding to $N+1$ points for $\theta$ and $\varphi$ at the Chebyshev nodes. The planes $(x{\rm O}y)$, $(y{\rm O}z)$, and $(z{\rm O}x)$ were all assumed to be planes of symmetry\footnote{The plane $(x{\rm O}y)$ is necessarily a plane of symmetry due to the Lichtenstein theorem.}, so that all calculations were performed in one-eighth of the fluid mass alone, which reduced the computation time. In this section, we benchmark BALEINES with analytical solutions and other numerical codes.

\begin{table*}[ht]
  \centering
  \caption{Comparison of numerical solutions from BALEINES with analytical solutions from the Maclaurin-Jacobi sequence for several resolutions.}
  \begin{tabular}{lrrrrrrrr}
        \hline
        &  \multicolumn{2}{c}{Analytical solutions} & \multicolumn{2}{c}{$N=8$} & \multicolumn{2}{c}{$N=16$} & \multicolumn{2}{c}{$N=32$}  \\
        \multicolumn{1}{c}{$c/a$} & \multicolumn{1}{c}{$b/a$} & \multicolumn{1}{c}{$\hat\varOmega^2$} & \multicolumn{1}{c}{$b/a$}  & \multicolumn{1}{c}{$\hat\varOmega^2$} & \multicolumn{1}{c}{$b/a$}  & \multicolumn{1}{c}{$\hat\varOmega^2$} & \multicolumn{1}{c}{$b/a$}  & \multicolumn{1}{c}{$\hat\varOmega^2$}\\\hline
        $0.95$ & $1.0000$ & $1.6563\times10^{-1}$ & $1.0000$ & $1.6568\times10^{-1}$ & $1.0000$ & $1.6566\times10^{-1}$ & $1.0000$ & $1.6563\times10^{-1}$ \\
        $0.90$ & $1.0000$ & $3.2689\times10^{-1}$ & $1.0000$ & $3.2699\times10^{-1}$ & $1.0000$ & $3.2690\times10^{-1}$ & $1.0000$ & $3.2689\times10^{-1}$ \\
        $0.85$ & $1.0000$ & $4.8288\times10^{-1}$ & $1.0000$ & $4.8303\times10^{-1}$ & $1.0000$ & $4.8289\times10^{-1}$ & $1.0000$ & $4.8288\times10^{-1}$ \\
        $0.80$ & $1.0000$ & $6.3256\times10^{-1}$ & $1.0000$ & $6.3276\times10^{-1}$ & $1.0000$ & $6.3259\times10^{-1}$ & $1.0000$ & $6.3257\times10^{-1}$ \\
        $0.75$ & $1.0000$ & $7.7473\times10^{-1}$ & $1.0000$ & $7.7497\times10^{-1}$ & $1.0000$ & $7.7476\times10^{-1}$ & $1.0000$ & $7.7474\times10^{-1}$ \\
        $0.70$ & $1.0000$ & $9.0799\times10^{-1}$ & $1.0000$ & $9.0826\times10^{-1}$ & $1.0000$ & $9.0802\times10^{-1}$ & $1.0000$ & $9.0799\times10^{-1}$ \\
        $0.65$ & $1.0000$ & $1.0307\mkern+53mu$ & $1.0000$ & $1.0310\mkern+53mu$ & $1.0000$ & $1.0307\mkern+53mu$ & $1.0000$ & $1.0307\mkern+53mu$ \\
        $0.60$ & $1.0000$ & $1.1409\mkern+53mu$ & $1.0000$ & $1.1412\mkern+53mu$ & $1.0000$ & $1.1409\mkern+53mu$ & $1.0000$ & $1.1409\mkern+53mu$ \\ 
        $0.55$ & $0.8943$ & $1.1698\mkern+53mu$ & $0.8954$ & $1.1711\mkern+53mu$ & $0.8944$ & $1.1700\mkern+53mu$ & $0.8943$ & $1.1698\mkern+53mu$ \\
        $0.50$ & $0.7545$ & $1.1389\mkern+53mu$ & $0.7553$ & $1.1404\mkern+53mu$ & $0.7546$ & $1.1391\mkern+53mu$ & $0.7545$ & $1.1389\mkern+53mu$ \\
        $0.45$ & $0.6351$ & $1.0830\mkern+53mu$ & $0.6358$ & $1.0848\mkern+53mu$ & $0.6352$ & $1.0832\mkern+53mu$ & $0.6351$ & $1.0830\mkern+53mu$ \\
        $0.40$ & $0.5316$ & $1.0038\mkern+53mu$ & $0.5322$ & $1.0060\mkern+53mu$ & $0.5317$ & $1.0041\mkern+53mu$ & $0.5316$ & $1.0038\mkern+53mu$ \\
        $0.35$ & $0.4406$ & $9.0329\times10^{-1}$ & $0.4410$ & $9.0588\times10^{-1}$ & $0.4407$ & $9.0362\times10^{-1}$ & $0.4406$ & $9.0333\times10^{-1}$ \\
        $0.30$ & $0.3596$ & $7.8384\times10^{-1}$ & $0.3598$ & $7.8634\times10^{-1}$ & $0.3597$ & $7.8424\times10^{-1}$ & $0.3597$ & $7.8389\times10^{-1}$ \\
        $0.25$ & $0.2868$ & $6.4846\times10^{-1}$ & $0.2845$ & $6.3067\times10^{-1}$ & $0.2850$ & $6.3323\times10^{-1}$ & $0.2849$ & $6.3252\times10^{-1}$ \\
        $0.20$ & $0.2206$ & $5.0129\times10^{-1}$ & $0.2173$ & $4.7560\times10^{-1}$ & $0.2172$ & $4.7714\times10^{-1}$ & $0.2172$ & $4.7680\times10^{-1}$ \\
        $0.15$ & $0.1598$ & $3.4835\times10^{-1}$ & $0.1591$ & $3.9550\times10^{-1}$ & $0.1591$ & $3.9023\times10^{-1}$ & $0.1591$ & $3.9027\times10^{-1}$ \\
        $0.10$ & $0.1035$ & $1.9904\times10^{-1}$ & $0.1044$ & $3.3702\times10^{-1}$ & $0.1048$ & $3.4539\times10^{-1}$ & $0.1048$ & $3.4553\times10^{-1}$ \\
        $0.05$ & $0.0506$ & $6.9914\times10^{-2}$ & $0.0520$ & $2.9461\times10^{-1}$ & $0.0521$ & $3.2992\times10^{-1}$ & $0.0520$ & $3.3096\times10^{-1}$ \\
        \hline
  \end{tabular}
  \label{tab:mljd}
\end{table*}

\subsection{Reproduction of the Maclaurin, Jacobi, and dumbbell sequences}

A first test for BALEINES is the comparison of its solutions with the analytical Maclaurin and Jacobi sequences. For this test, we set $L=1$, and the only input was then the meridian axis ratio $c/a$. The results are reported in Table \ref{tab:mljd}, where the values of $b/a$ and $\hat{\varOmega}^2$ obtained with BALEINES are compared with the analytical solution. The deviation from the references is consistently on the same order of magnitude for $c/a\geq0.30$. Doubling the resolution decreases the error by almost an order of magnitude, from $\sim 2\times 10^{-4}$ for $N=8$ to $\sim 3\times10^{-6}$ for $N=32$. For Clenshaw-Curtis quadratures, this is underwhelming, but is clearly due to the nonderivable character of the kernel (see the discussion in Sect. \ref{ssec:psinum}).

\begin{figure}[ht]
  \centering
  \includegraphics[width=0.85\linewidth]{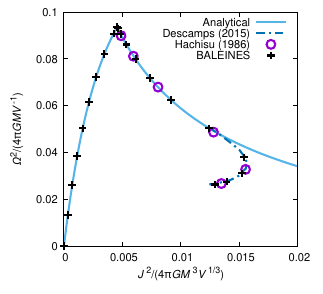}
  \caption{Comparison of numerical solutions from BALEINES ($N=16$) with analytical solutions from the Maclaurin-Jacobi sequence and numerical solutions reported by \citet{hachisu86III} and \citet{descamps15}.}
  \label{fig:mljd}
\end{figure}

For $c/a\leq 0.25$, the deviation of the analytical solution is greater than the aforementioned estimate. This is also visible in Fig. \ref{fig:mljd}, where the solutions are reported on a angular momentum squared - rotation rate squared diagram. These equilibria, where the deviation from ellipsoids is significant, belong to the dumbbell sequence \citep{ehs82} that branches off the Jacobi sequence at $(b/a,c/a)\approx(0.305,0.265)$. This is confirmed by the comparison with the solutions obtained by \citet{hachisu86III} and \citet{descamps15} in Fig. \ref{fig:mljd}. 

\begin{figure}[ht]
  \centering
  \includegraphics[width=0.85\linewidth]{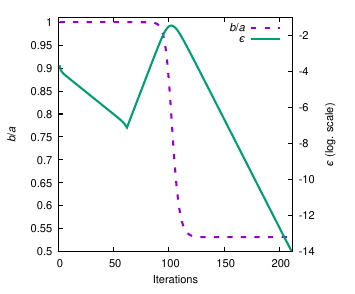}
  \caption{Evolution of the convergence indicator, $\epsilon$, and equatorial axis ratio, $b/a$, during the cycle for an initial ellipsoid with $c/a=0.40$, $b/a=1.00$, and $N=16$.}
  \label{fig:bounce}
\end{figure}

It is interesting to note that the Jacobi sequence is unstable beyond this point, whereas the dumbbell is stable. In the same manner, the Maclaurin sequence is unstable beyond Meyer's bifurcation point, that is, $(b/a,c/a)\approx (1,0.583)$, and the Jacobi sequence is stable between these two points. Any attempt to obtain an axisymmetric configuration beyond Meyer's point first converged to a Maclaurin spheroid to a point where the convergence indicator, $\epsilon$, reached $\sim10^{-7}$ before it bounced back up and converged to the Jacobi ellipsoid with the same $c/a$ (see Fig. \ref{fig:bounce}). This behavior is understandable by noting that the numerical precision of the quadratures acts as a perturbation of the free surface of the fluid, leading the SCF-cycle to naturally bifurcate from the unstable to the stable state. The same phenomenon occurred when we tried to obtain a Jacobi ellipsoid with $c/a \lesssim 0.265$: BALEINES naturally converged toward the corresponding dumbbell instead. To obtain these solutions, the convergence threshold can be set higher (typically, $\epsilon_0=10^{-6}$ in the example in Fig. \ref{fig:bounce}) to stop the cycle before the bifurcation occurs. 

\subsection{Comparison with the {\tt kyushu} code}

\begin{figure}[ht]
  \centering
  \includegraphics[width=0.9\linewidth]{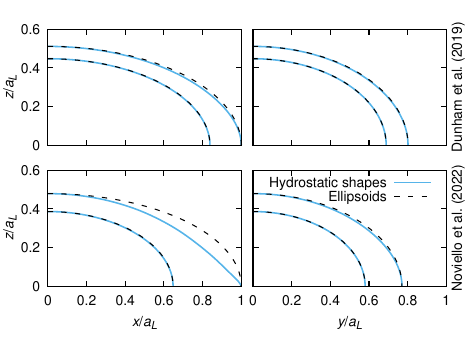}
  \caption{Equilibrium figures of the free surface and the core-mantle boundary obtained with BALEINES for the configurations reported in Tables \ref{tab:kyushu1} (top) and \ref{tab:kyushu2} (bottom) in the $(x{\rm O}z)$ (left) and $(y{\rm O}z)$ (right) planes. The equivalent ellipsoids are also shown as dashed lines.}
  \label{fig:kyushu}
\end{figure}

As a second test, we compared two-layer solutions obtained with the {\tt kyushu} code \citep{ddp19,noviello22}. This code, based on the algorithm reported by \citet{hachisu86III}, was written to study the hydrostatic structure of a planetoid with a silicate core and an icy mantle. The inputs of these models are the length of the two equatorial semiaxes, $a_2$ and $b_2$, the mantle mass density, $\rho_2$, the mass of the object, $M$ and its rotation rate, $\varOmega$. The latter two quantities are not used in BALEINES for the convergence, but can be used after convergence to recover all physical quantities. We have
\begin{equation}
  \left\{
    \begin{aligned}
      \rho_2 &= \frac{\varOmega^2}{G\hat\varOmega^2},\\
      a_2 &= \left(\frac{GM}{\varOmega^2}\frac{\hat\varOmega^2}{\hat{M}}\right)^{1/3},
    \end{aligned}
  \right.
\end{equation}
and all other quantities are readily deduced by rescaling. We chose to use the configuration preferred by \citet{ddp19} for Haumea and a configuration of the tables of \citet{noviello22} as references. The comparisons between BALEINES and {\tt kyushu} are reported in Tables \ref{tab:kyushu1} and \ref{tab:kyushu2}, respectively. The resolution used in both cases was $N=16$. Tests were made with $N=32$ as well, but the results were the same within the number of digits reported for the reference solutions. The agreement between the two methods is satisfying, as the deviation is about a few $10^{-3}$ in Table \ref{tab:kyushu1} and below $1~\%$ in Table \ref{tab:kyushu2}, which is the numerical precision stated by \citet{ddp19} and \citet{noviello22} for {\tt kyushu}. The largest deviations are found on the mean density, with errors between $1$ and $10~\%$. This is because in {\tt kyushu}, the mean density was computed as if the interfaces between layers and the free surfaces were ellipsoids. As shown in Fig. \ref{fig:kyushu}, the volume is overestimated because of this approximation, meaning that the mean density, $\bar{\rho}$, is underestimated. In the case of the configuration reported by \citet{ddp19}, the deviation is weak (within $3~\%$), so the estimate of $\bar{\rho}$ is still correct; this is not the case for the configuration of \citet{noviello22}, however.

\begin{table}[ht]
  \centering
  \caption{Comparison between the inputs and outputs of {\tt kyushu} and BALEINES for the model of Haumea preferred by \citet{ddp19}}
  \begin{tabular}{lrr}
    & \multicolumn{1}{c}{\tt kyushu} & \multicolumn{1}{c}{BALEINES}\\\hline
    $M~(10^{21}~{\rm kg})$ & $\leftarrow 4.006$ & $\leftarrow 4.006$ \\
    $2\uppi/\varOmega~({\rm h})$ & $\leftarrow 3.91531$ & $\leftarrow 3.91531$\\
    $c_2/a_2$ & $0.51143$ & $\leftarrow 0.51143$\\
    $\rho_1/\rho_2$ & $2.90988$ & $\leftarrow 2.90988$ \\
    $c_1/c_2$ & $0.87523$ & $\leftarrow 0.87523$ \\
    $a_2~\mathrm{(km)}$ & $\leftarrow 1050$ & $1051$ \\
    $b_2~\mathrm{(km)}$ & $\leftarrow 840$ & $843$ \\
    $c_2~\mathrm{(km)}$ & $537$ & $537$ \\
    $a_1~\mathrm{(km)}$ & $883$ & $881$ \\
    $b_1~\mathrm{(km)}$ & $723$ & $724$ \\
    $c_1~\mathrm{(km)}$ & $470$ & $470$ \\
    $\rho_1~\mathrm{(kg\cdot m^{-3})}$ & $2680$ & $2683$ \\
    $\rho_2~\mathrm{(kg\cdot m^{-3})}$ & $\leftarrow 921$ & $922$ \\
    $\bar\rho~\mathrm{(kg\cdot m^{-3})}$ & $2018$ & $2053$ \\
    $P_{1}~{\rm (MPa)}$ & $30.4$ & $30.8$ \\\hline
    $N_r\times N_\theta\times N_\varphi$ & $391\times33\times33$ & $2\times17\times17$ \\\hline
  \end{tabular}
  \tablefoot{``$\leftarrow$'' indicates the input quantities. $N_r$, $N_\theta$ and $N_\varphi$ are the number of nodes in the radial, polar, and azimuthal directions, respectively. }
  \label{tab:kyushu1}
\end{table}

\begin{table}[ht]
  \centering
  \caption{Comparison between the inputs and outputs of {\tt kyushu} and BALEINES for model 16 in Table 2 of \citet{noviello22}}
  \begin{tabular}{lrr}
    & \multicolumn{1}{c}{\tt kyushu} & \multicolumn{1}{c}{BALEINES}\\\hline
    $M~(10^{21}~{\rm kg})$ & $\leftarrow 4.126$ & $\leftarrow 4.126$ \\
    $2\uppi/\varOmega~({\rm h})$ & $\leftarrow 3.393$ & $\leftarrow 3.393$\\
    $c_2/a_2$ & $0.47853$ & $\leftarrow 0.47853$\\
    $\rho_1/\rho_2$ & $4.31705$ & $\leftarrow 4.31705$ \\
    $c_1/c_2$ & $0.80636$ & $\leftarrow 0.80636$ \\
    $a_2~\mathrm{(km)}$ & $\leftarrow 1104$ & $1102$ \\
    $b_2~\mathrm{(km)}$ & $\leftarrow 840$ & $849$ \\
    $c_2~\mathrm{(km)}$ & $528.3$ & $527.5$ \\
    $a_1~\mathrm{(km)}$ & $714$ & $714$ \\
    $b_1~\mathrm{(km)}$ & $636$ & $639$ \\
    $c_1~\mathrm{(km)}$ & $426$ & $425$ \\
    $\rho_1~\mathrm{(kg\cdot m^{-3})}$ & $3976$ & $3982$ \\
    $\rho_2~\mathrm{(kg\cdot m^{-3})}$ & $\leftarrow 921$ & $922$ \\
    $\bar\rho~\mathrm{(kg\cdot m^{-3})}$ & $2011$ & $2289$ \\
    $P_{1}~{\rm (MPa)}$ & $53$ & $58$ \\\hline
    $N_r\times N_\theta\times N_\varphi$ & $201\times17\times17$ & $2\times17\times17$ \\\hline
  \end{tabular}
  \tablefoot{``$\leftarrow$'' indicates the input quantities. $N_r$, $N_\theta$ and $N_\varphi$ are the number of nodes in the radial, polar, and azimuthal directions, respectively.}
  \label{tab:kyushu2}
\end{table}

\section{Equilibrium sequences for differentiated bodies}\label{sec:examples}
\subsection{Evolution of a two-layer system with increasing polar flattening}\label{ssec:sequence}

The objective of this section is to examine how a second layer alters the homogeneous sequence shown in Fig.  \ref{fig:mljd}. These studies were conducted in the case of polytropic fluids \citep[e.g.,][]{hachisu86III}, that is, fluids in which the pressure and the mass density are related by $P=K\rho^{1+1/n}$, $K$ and $n$ being positive constants; $n$ is the natural parameter of the sequences for polytropes. We chose to impose the core volume fraction (CVF), that is, the ratio of the volume of the core and the total volume of the object (see Sect. \ref{ssec:om2cte}). Holding this fraction fixed allowed us to keep the mass of both components constant along the sequence, which is interesting as it may be interpreted as the evolution of a two-layer system as its polar flattening increases. However, imposing the CVF has the consequence to slow down the convergence of the SCF-cycle because no point on the core-mantle boundary is fixed. 

\begin{figure}[ht]
  \centering
  \includegraphics[width=0.85\linewidth]{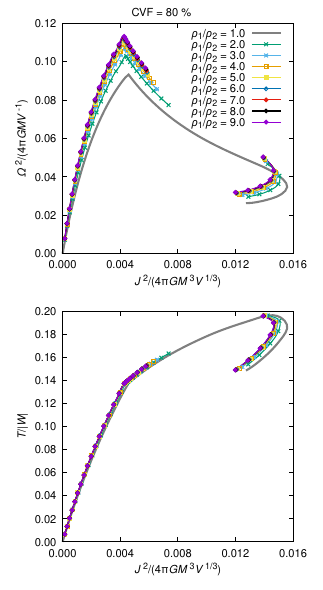}
  \caption{Normalized rotation rate (top) and stability indicator (bottom) against normalized angular momentum for several equilibrium sequences of two-layer systems with a CVF of $80~\%$. The homogeneous stable sequence of the equilibrium (Maclaurin-Jacobi-Dumbbell) is plotted in gray for reference.}
  \label{fig:seqvol_800}
\end{figure}

\begin{figure}[ht]
  \centering
  \includegraphics[width=0.85\linewidth]{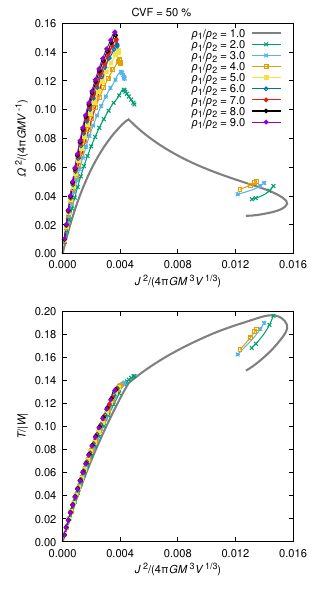}
  \caption{Same as Fig. \ref{fig:seqvol_800}, but for a CVF of $50~\%$.}
  \label{fig:seqvol_500}
\end{figure}

The sequences are shown in Figs. \ref{fig:seqvol_800} and \ref{fig:seqvol_500} for core volume fractions of $80~\%$ and $50~\%$, respectively, with the rotation rate squared and the stability indicator as functions of the angular momentum squared. The curves obtained for differentiated bodies resemble their polytropic counterpart \citep[see][]{hachisu86III}, with a shift toward higher rotations and smaller angular momentum as the body approaches homogeneity, and with open sequences, that is, there is no longer a triaxial-dumbbell bifurcation as the Jacobi-like sequence ends prematurely due to mass shedding. However, it is interesting to note that $T/|W|$ is nearly unchanged by the presence of a deeper layer, as only the shift toward lower values of $J^2$ is clearly visible. 

\subsection{Meyer bifurcation for two-layer systems. Application to Quaoar}\label{ssec:bifurcation}

As mentioned in the introduction and in Sect. \ref{sec:tests}, the axisymmetric-triaxial bifurcation occurs for $c/a\approx0.58272$. The bifurcation in heterogeneous bodies was studied by \citet{j64} and \citet{vandervoort80} in the case of polytropic fluids. The former showed that no bifurcation is possible only for $n\lesssim0.808$ because of mass shedding, while the latter showed that Meyer's point is shifted toward faster rotations as $n$ increases. 

The question of the bifurcation of differentiated bodies is particularly interesting for Quaoar. As mentioned in the introduction, \citet{kiss2024} recently proposed thermophysical models of the surface of this object, and the model that matched the observed thermal light curve best was a triaxial ellipsoid with $b/a \approx 0.840$ and $c/a \approx 0.724$. However, the obtained axis ratios are inconsistent with a Jacobi ellipsoid, as shown in Fig. \ref{fig:hauqua_vs_jac}, and are very far from the homogeneous sequence. For Quaoar to be hydrostatic, a sequence that bifurcates from axisymmetry at $c/a > 0.724$ would be necessary. Thus, studying the position of the bifurcation for differentiated systems might help us to conclude on the hydrostaticity of the model of \citet{kiss2024}. 

For this study, we used BALEINES in a mode in which we held the points along the equatorial minor semiaxis fixed instead of the polar semiaxis (see Sect. \ref{ssec:om2cte}). We then ran BALEINES with $b_2/a_2 = 0.999$, $b_1/b_2 \in [0.05,0.95]$ with step $0.05$, and $\rho_1/\rho_2 \in [2,9]$ with step $1$.

\begin{figure}[ht]
  \centering
  \includegraphics[width=0.85\linewidth]{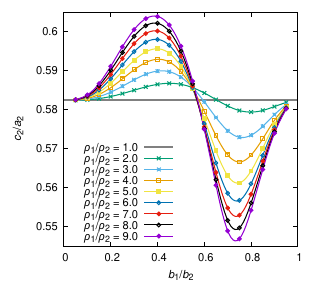}
  \caption{Polar to major equatorial semiaxes of the free surface, $c_2/a_2$, corresponding to the axisymmetric-triaxial bifurcation as a function of $b_1/b_2$ for several values of $\rho_1/\rho_2$. The Jacobi ellipsoid with $b/a=0.999$, which has $c/a\approx0.5824$, is plotted as well for reference. Cubic splines were used to draw the plain lines.}
  \label{fig:bal_bifurc}
\end{figure}

Figure \ref{fig:bal_bifurc} shows that bifurcation is shifted toward slower rotations (i.e., higher $c_2/a_2$) for small cores ($b_1/b_2\lesssim 0.55-0.65$) and toward faster rotations (i.e., lower $c_2/a_2$) for large cores. As $b_1/b_2$ tends to either $0$ or $1$, the fluid is closer to homogeneity and the bifurcation then tends to Meyer's point as expected. However, the value of $c_2/a_2$ at the bifurcation deviates by less than $10~\%$ from Meyer's point, with a maximum value of $0.604$. This is still clearly not close enough to the required (minimum) value of $0.724$. We note that $\rho_1/\rho_2=9$ is already unrealistic for a trans-Neptunian object: this value is higher than the density ratio of iron and water. Even considering an absurdly high value of $\rho_1/\rho_2=35$, the maximum value of $c_2/a_2$ only reaches $0.626$. We can then conclude that the triaxial shape derived by \citet{kiss2024} is inconsistent with an object at hydrostatic equilibrium. The study of the bifurcation point is clearly not the only method to obtain this conclusion \citep[see, e.g.,][]{kondra23}.

\section{Discussion}\label{sec:discussion}

We reported a new numerical code, BALEINES, for studying the interior structure of a rotating triaxial mass composed of homogeneous layers in hydrostatic equilibrium. The efficiency of the method was ensured by an adaptive meshing that optimizes the number of radial points needed, that is, only one per layer. The gravitational potential was obtained via a sum of surface integrals. These surface integrals do not present the point-mass divergence of the classical triple integral, which facilitates the numerical integration. The performances of the code agree very well with analytical solutions and the {\tt kyushu} code \citep{ddp19,noviello22}, which was used to study the interior of the dwarf planet Haumea. We then used BALEINES to study the axisymmetric-triaxial bifurcation of two-layer systems and showed that the thermophysical model of the surface preferred by \citet{kiss2024} is inconsistent with a mass at hydrostatic equilibrium. 

A short-term perspective clearly is to use BALEINES to study whether Haumea is a differentiated body at hydrostatic equilibrium. The adaptive meshing would allow us to scan the parameter space in the two-layer case in a more refined way to search for hydrostatic solutions close to observations or previous models. If the two-layer models were encouraging, it would be a thrilling possibility to explore the three-layer parameter space to constrain the internal structure of Haumea based on its triaxial shape alone.

An external potential in the form of a multipolar expansion, for example, could be introduced in $\varPsi$ to study the effect of tides on the figure. This might be of interest for modeling the hydrostatic shape of a differentiated satellite in the context of the JUICE mission \citep{JUICE} toward the moons of Jupiter. In any case, the system has to be tidally locked for the problem to be still independent of time in the corotating frame. This condition is mandatory for using the Bernoulli equation. We might also consider a whole fluid instead of a prescribed potential and study the interaction between the two masses.

\begin{acknowledgements} 
    We are grateful to N. Rambaux and F. Chambat for stimulating discussions. We thank the anonymous referee for bibliographic inputs and comments to improve the paper.
\end{acknowledgements}

\bibliographystyle{aa}
\bibliography{baleines}

\begin{appendix}

\section{Gravitational energy of a homogeneous body}\label{app:w_proof}

The gravitational energy of a self-gravitating body reads
\begin{equation}
    W = \frac12\iiint_{\cal V} \du^3\vec{r}\, \rho(\vec{r}) \varPsi(\vec{r}),
\end{equation}
which can be also written as
\begin{equation}
  W = -\iiint_{\cal V} \du^3\vec{r}\, \rho(\vec{r})\vec{r}\cdot\del\varPsi.
\end{equation}
First, we assume that the body is fully homogeneous and we integrate by parts:
\begin{equation}
  W = -\rho \iiint_{\cal V} \du^3\vec{r}\,\del\cdot(\vec{r}\varPsi) + 3\rho\iiint_{\cal V} \du^3\vec{r}\,\varPsi(\vec{r}).
\end{equation}
The second term of the right-hand side is clearly $6W$. So, using the divergence theorem, we finally obtain
\begin{equation}\label{eq:wenergy_2d}
  W = \frac15\rho\oiint_{{\cal S}} \du^2\vec{r}\,(\vec{n}\cdot\vec{r})\varPsi(\vec{r}).
\end{equation}
This expression is attractive in our framework, as it only requires the values of the gravitational potential on the surface of the mass. 

\end{appendix}

\end{document}